\documentclass[12pt]{article}  
\usepackage{times} 
\usepackage{fleqn} 
\usepackage{amssymb,bm} 
\usepackage{epsfig,graphicx,color}  
\usepackage{import} 

\begin{document}

\addtolength{\baselineskip}{0.20\baselineskip}

\vspace{48pt}

\centerline{\Huge Three dimensional four-fermion models -}
\bigskip
\centerline{\Huge A Monte Carlo study}
\vspace{28pt}

\centerline{\bf Stavros Christofi}
                                                                                                                                      
\medskip
\smallskip                                                                                                                                      
\centerline{ {\sl Frederick Institute of Technology, 1303 Nicosia, Cyprus.}}
\bigskip

\centerline{\bf Costas Strouthos}
\medskip
\centerline{ {\sl Harvard-MIT (HST) Martinos Center for Biomedical Imaging,}}
\centerline{\sl Massachusetts General Hospital, Harvard Medical School, Charlestown, MA 02129}

\vspace{24pt}
                                                                                                                                      
                                                                                                                                      
\centerline{{\bf Abstract}}
                                                                                                                                      
\noindent
{\narrower
We present results from numerical simulations of three different $3d$ four-fermion models that exhibit $Z_2$,
$U(1)$, and $SU(2) \times SU(2)$ chiral symmetries, respectively. We performed the simulations by using the hybrid
Monte Carlo algorithm.  We employed finite size scaling methods
on lattices ranging from $8^3$ to $40^3$
to study the properties of the second order chiral phase transition in each model.
The corresponding critical coupling defines an ultraviolet fixed point of the renormalization
group.
In our high precision simulations, we detected next-to-leading order corrections for various critical exponents
and we found them to be in good agreement with existing analytical large-$N_f$ calculations.
}

\bigskip
\noindent

\vfill
\newpage

\section{Introduction}
The 3d four-fermion models are among the simplest relativistic quantum field theories
of interacting fermions. There are several motivations for studying such models.
Dynamical breaking of chiral symmetry occurs at strong enough interaction coupling $g^2_c$.
The chirally broken phase is separated
from the chirally symmetric phase by a second order phase transition at the critical coupling.
Even though these models are
not perturbatively renormalizable, it has been shown that the $1/N_f$ expansion
about the fixed point $g^2_c$ is exactly renormalizable \cite{rosenstein.1}.
In addition, four-fermion models are ideal laboratories for
studying continuum phase transitions in the presence of massless fermions. Hence, they define new universality
classes that are quantitatively different from the ferromagnetic phase transitions in bosonic $O(N)$ Heisenberg
spin models. Furthermore, in the framework of $1/N_f$ expansion \cite{justin}, it has been shown that the universality class
of the $d$-dimensional four-fermion models, where $d$ is between two and four, is the same as the universality class of the Higgs-Yukawa model
with the same chiral symmetry.
Understanding the properties of the
continuum phase transition, which separates the chirally symmetric from the chirally broken phase, requires
non-perturbative techniques such as the large-$N_f$ expansion \cite{rosenstein.1, rosenstein.2, hands, derkachov, gracey.z2a,
gracey.z2b, gracey.su2.u1, gracey.u1b, gracey.su2b},
exact renormalization group equations \cite{wetterich.1, wetterich.2},
and lattice Monte Carlo simulations \cite{hands,karkkainen,focht}.

Given that $3d$ four-fermion models incorporate
certain important features of QCD, they have been used recently as model field theories to study
the properties of the strong interaction at non-zero temperature and non-zero quark number density
\cite{general}.
In addition, there may be applications of four-fermion  models to high-$T_c$ 
superconductivity \cite{sharkar}, for instance in describing non-Fermi liquid behavior in the 
normal phase \cite{aitchison}. More recently, it was proposed that the Hubbard model on a 
honeycomb lattice relevant to the newly discovered graphene sheets, has a transition described by the three-dimensional
$Z_2$-symmetric four-fermion model \cite{herbut}.

In this paper, we study numerically the critical properties of the $3d$ four-fermion models that exhibit the three
different $Z_2$, abelian $U(1)$, and non-abelian $SU(2)\times SU(2)$
chiral symmetries.
In our simulations, we fixed the number of fermion flavors to $N_f=4$.
Our simulations are the first accurate finite size scaling (FSS) studies of the $U(1)$ and $SU(2) \times SU(2)$ models
that allow us to detect next-to-leading order corrections on the values of the critical exponents and to compare them
with existing analytical large-$N_f$ predictions \cite{derkachov,gracey.z2a,gracey.z2b,gracey.su2.u1,gracey.u1b, gracey.su2b}. 
Our results from the $Z_2$ model simulations are also in good agreement with
existing large-$N_f$ predictions as are other accurate Monte Carlo results with $N_f=2$ \cite{karkkainen}.

\section{Models and Observables}
In this section, we introduce the three different versions of the model we shall be dealing with in the
bulk of this paper and the observables used to measure the critical exponents of the continuous
phase transitions. In the literature, the models are often called the Gross-Neveu models
and their continuum space-time lagrangians (we work in Euclidean space) are as follows:
\begin{equation}
{\cal L}_A =
\bar\Psi_i
(\partial{\!\!\! /}\,+m)\Psi_i-{g^2\over{2N_f}}(\bar\Psi_i\Psi_i)^2
\end{equation}
\begin{equation}
{\cal L}_B = \bar{\Psi}_i(\partial\hskip -.5em / + m) \Psi_i
- \frac{g^2}{2 N_f} [(\bar{\Psi}_i \Psi_i)^2 - (\bar{\Psi}_i \gamma_5 \Psi_i)^2].
\end{equation}
\begin{equation}
{\cal L}_C =\bar\Psi_i({\partial\!\!\!/\,}+m)\Psi_i -{g^2\over2}
\left[(\bar{\Psi}_i \Psi_i)^2-(\bar{\Psi}_i \gamma_5\vec\tau\Psi_i)^2\right].
\end{equation}
We treat $\Psi_i$, $\bar{\Psi}_i$  as four-component Dirac spinors and the index $i$ runs over $N_f$
fermion species. It can be easily shown that in the chiral limit $m \to 0$, ${\cal L}_A$ has a
$Z_2$, ${\cal L}_B$ a $U(1)$, and ${\cal L}_C$ an
$SU(2) \times SU(2)$ chiral symmetry.

For analytical and computational purposes, it
is useful to introduce auxiliary fields $\sigma$ and $\pi_i$. Hence, the bosonized lagrangians become
quadratic in $\Psi_i$:
\begin{equation}
{\cal L}_A =
\bar\Psi_i
(\partial{\!\!\! /}\,+m +\sigma)\Psi_i+{N_f\over{2g^2}}
\sigma^2 .
\end{equation}
\begin{equation}
{\cal L}_B= \bar{\Psi}_i(\partial\hskip -.5em / + m + \sigma + i \gamma_5 \pi)\Psi_i
+ \frac{N_f}{2 g^{2}} (\sigma^{2}+ \pi^2)
\end{equation}
\begin{equation}
{\cal L}_C  =  \overline{\Psi}_i\left(\partial\hskip -.5em / + m + \sigma+i\gamma_5
\vec\pi \cdot \vec\tau\right)\Psi_i
+\frac{N_c}{2g^2}\left(\sigma^2+\vec\pi \cdot \vec\pi\right).
\end{equation}
For sufficiently strong coupling $g^2 > g^2_c$ the models exhibit spontaneous symmetry breaking
implying dynamical generation of a fermion mass. The pion fields $\pi_i$ become the associated Goldstone bosons.

We used the staggered fermion discretization with the auxiliary fields living on the dual lattice sites
to formulate the models in their bosonized form on the lattice.
For each case, we used the hybrid Monte Carlo algorithm with $N_f=4$ fermion flavors
to perform numerical simulations exactly. The Monte Carlo procedure was optimized by choosing the microcanonical 
trajectory length at random from a Poisson distribution with mean value equal to 1.0. This method of optimization 
which guarantees ergodicity was found to decrease the autocorrelations in the data significantly \cite{wang}.  
Details concerning the lattice actions and the numerical algorithm can be found
in \cite{hands, hands.u1, hands.su2}. 

We work in the chiral limit to study the chiral phase transition of the models. Hence, we choose
not to introduce a bare quark mass into the lattice action. Without the benefit of this interaction,
the direction of symmetry breaking changes over the course of the simulation such that
$\Sigma \equiv \frac{1}{V} \sum_x \sigma(x)$ and $\Pi_i \equiv \frac{1}{V} \sum_x \pi_i(x)$
average to zero over the ensemble.
It is in this way that the absence of spontaneous symmetry breaking on a finite lattice is enforced. Another
option is to introduce an effective order parameter $\Phi$ equal to the magnitude of the vector
$\vec{\Phi}\equiv (\Sigma, \vec{\Pi})$. In the thermodynamic limit, $\langle \Phi \rangle$ is equal to the true
order parameter $\langle \sigma \rangle$ extrapolated to zero quark mass.

We employ the finite size scaling (FSS) method \cite{barber}, a well-established tool,
to study the critical behavior of the model on lattices available to us.
The correlation length $\xi$ on a finite lattice is limited by the size of the system and consequently
no true criticality can be observed.
The dependence of a given thermodynamic observable, $A$, on the
size $L$ of the box is singular. According to the FSS
hypothesis, in the large volume limit, $A$ is given by:
\begin{equation}\label{fssX}
A(t,L) = L^{\rho_A/\nu}Q_A(tL^{1/\nu}),
\end{equation}
where $t \equiv (\beta_c-\beta)/\beta_c$ is the reduced temperature,
$\nu$ is the exponent of the correlation length,
$Q_A$ is a scaling function that is not
singular at zero argument, and
$\rho_A$ is the critical exponent
for the quantity $A$.
Using eq.~(\ref{fssX}), one can determine such exponents
by measuring $A$ for different values of $L$.

In the large $L$ limit, the FSS scaling form of the effective
order parameter $\langle \Phi \rangle$ is given by
\begin{equation}
\langle \Phi \rangle =  L^{-\beta_m/\nu}f_{\sigma}(tL^{1/\nu}).
\label{eq:magn1}
\end{equation}

A standard method to measure the inverse critical coupling $\beta_c \equiv 1/g^2$ for a second order
transition is to compute the Binder cumulant $U_B(\beta,L)$ \cite{binder}, defined by
\begin{equation}
U_B \equiv 1-\frac{1}{3}\frac{\langle\Phi^4\rangle}{\langle\Phi^2\rangle^2},
\end{equation}
for various system sizes.
Near the critical coupling and on sufficiently large lattices, where subleading corrections from the finite lattice size $L$
are negligible, $U_B=f_B L(tL^{1/\nu})$. Therefore, at $\beta_c$, $U_B$ becomes independent of $L$.
Deviations from this relation can be explained by finite size confluent corrections.
The leading $L_1/L$ dependence in the deviation of the intersection point $\beta_*$ from the critical point
$\beta_c$ is estimated by Binder \cite{binder} as
\begin{equation}
\frac{1}{\beta_*(L)} = \frac{1}{\beta_c} + \frac{a}{{\ln}(L_1/L)}.
\label{eq:confl}
\end{equation}
In our analysis we chose $L$ to be the smallest lattice size $L=8$ and hence $L_1$ are the remaining lattice sizes.

For the general $O(n)$-symmetric models, it can be easily shown \cite{holm}
that as the lattice volume tends to infinity in the weak coupling limit,
Gaussian fluctuations around $\vec{\Phi}=0$ lead to $U_B \to 2(n-1)/3n$. For $n=1$ ($O(1) \equiv Z_2$ symmetry)
this gives a zero reference point, for $n=2$ ($O(2) \equiv U(1)$ symmetry) $U_B \to 1/3$, and for
$n=4$ ($O(4) \equiv SU(2) \times SU(2)$ symmetry) $U_B \to 1/2$. In the chirally broken phase $U_B \to 2/3$
for all $n$ in the thermodynamic limit.

Another quantity of interest is the susceptibility $\chi$ that is given, in the static limit of the
fluctuation-dissipation theorem, by
\begin{equation}
\chi = \lim_{L \to \infty} V[\langle \vec{\Phi}^2 \rangle - \langle \vec{\Phi} \rangle \cdot \langle \vec{\Phi} \rangle],
\end{equation}
where $V$ is the lattice volume.
For finite systems, the true order parameter $\langle \vec{\Phi} \rangle $ vanishes
and for $\beta \geq \beta_c$ the susceptibility is given by:
\begin{equation}
\chi = V\langle \Phi^2 \rangle.
\label{eq:chi1}
\end{equation}
This observable should scale at criticality like
\begin{equation}
\chi = L^{\gamma/\nu}f_{\chi}(tL^{1/\nu}).
\label{eq:chi.fss}
\end{equation}

Furthermore, the logarithmic derivatives of $\langle \Phi \rangle$ can give estimates for the critical exponent $\nu$.
It can be easily shown that
\begin{equation}
D \equiv \frac{\partial}{\partial \beta}  \ln \langle \Phi \rangle  =
\left[ \frac{\langle \Phi S_b \rangle}
{\langle \Phi \rangle} - \langle S_b \rangle \right],
\label{eq:derlog}
\end{equation}
where $S_b$ 
is the bosonic part of the lattice action that is multiplied by the coupling $\beta$.
$D$  has a scaling relation
\begin{equation}
D = L^{1/\nu}f_D(tL^{1/\nu}).
\label{eq:ld}
\end{equation}

We used the histogram reweighting method \cite{histo}
to perform our study most effectively. This
enabled us to calculate the observables in a region of couplings around the simulation coupling.
We utilized this technique efficiently by performing simulations at
slightly different couplings $\beta_i$ close to the critical coupling $\beta_c$.
We also employed the jacknife method to estimate the statistical errors on the various observables 
reliably. This method accounts for correlations in the raw data set.

\section{Results}

\begin{figure}[t!]
\bigskip\bigskip
\begin{center}
\epsfig{file=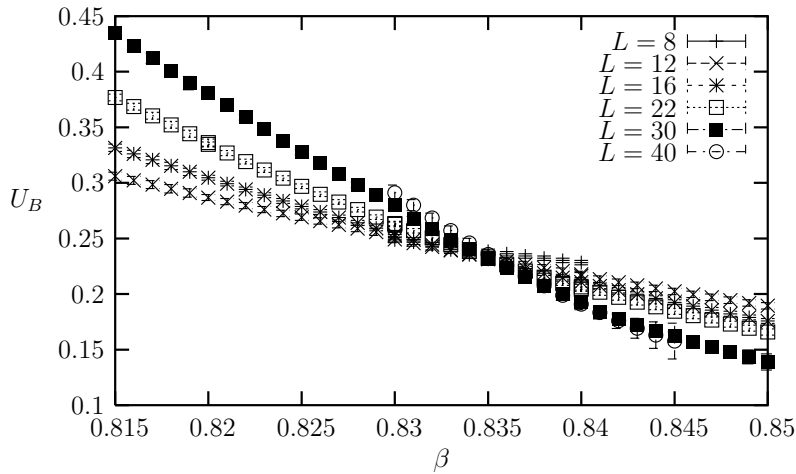, width=10.5cm}
\end{center}
\caption{Binder cumulant vs. $\beta$ for different lattice sizes; $Z_2$ model.}
\label{fig:bind_z2}
\end{figure}

In this section we present the results of the data analysis for the three different models.
In all three cases, the fermion species number is fixed at $N_f=4$.
An accurate determination of the critical exponents requires a precise determination of the 
critical coupling.
We calculated the critical couplings by using the Binder cumulant technique described in the previous section. 
For different lattice sizes, the curves $U_B=U_B(\beta)$ should intersect at $\beta=\beta_c$ up to finite size corrections that are
visible on the smaller lattices. 
We used the histogram reweighting method to obtain the values of $U_B$ versus $\beta$. We show these values for the $Z_2$
model in Fig.~\ref{fig:bind_z2}.
\begin{figure}[b!]
\bigskip\bigskip
\begin{center}
\epsfig{file=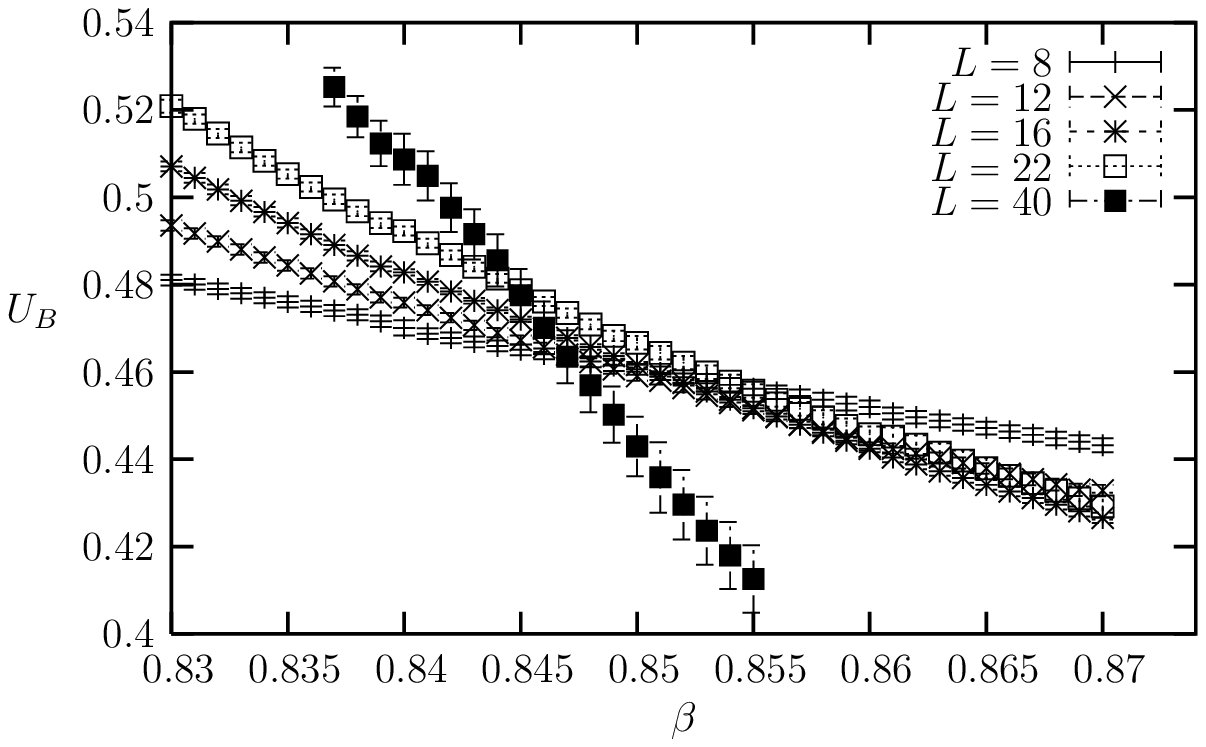, width=10.5cm}
\end{center}
\caption{Binder cumulant vs. $\beta$ for different lattice sizes; $U(1)$ model.}
\label{fig:bind_u1}
\end{figure}
\begin{figure}[t!]
\bigskip\bigskip
\begin{center}
\epsfig{file=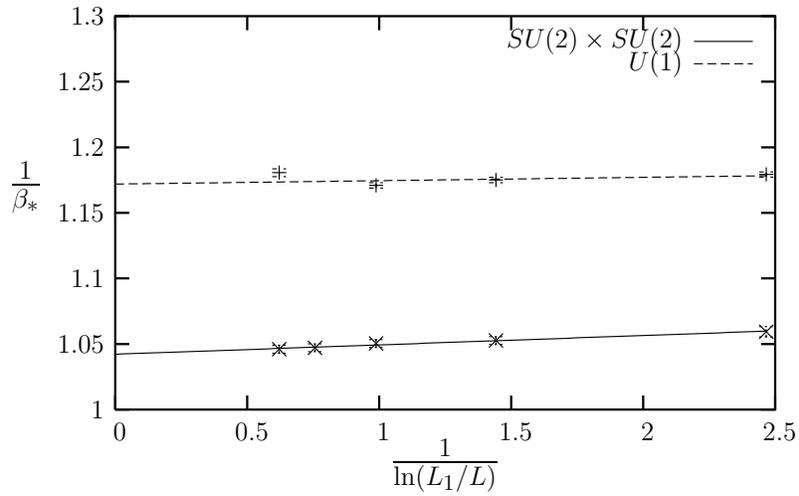, width=10.5cm}
\end{center}
\caption{The intersection of $U_{B}$$(L)$ and $U_{B}$$(L_1)$ for $L=8$ vs. $\ln(L_1/L)$.}
\label{fig:bind_betac}
\end{figure}
\begin{figure}[b!]
\bigskip\bigskip
\begin{center}
\epsfig{file=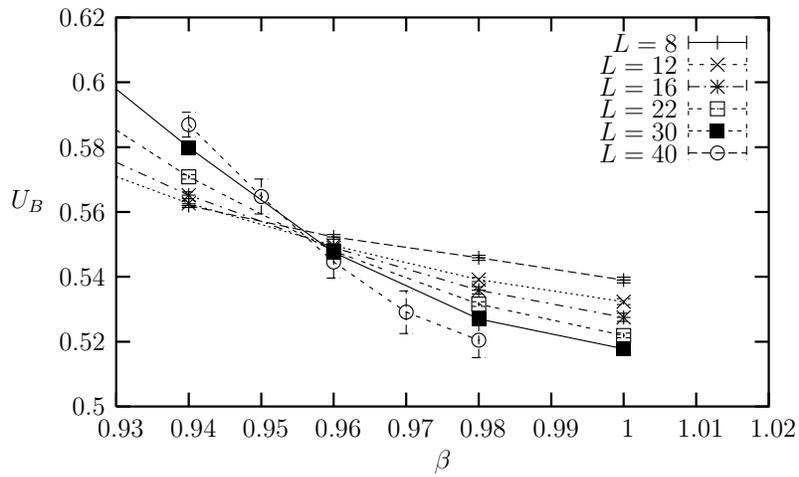, width=10.5cm}
\end{center}
\caption{Binder cumulant vs. $\beta$ for different lattice sizes; $SU(2) \times SU(2)$ model.}
\label{fig:bind_su2}
\end{figure}
\begin{figure}[t!]
\bigskip\bigskip
\begin{center}
\epsfig{file=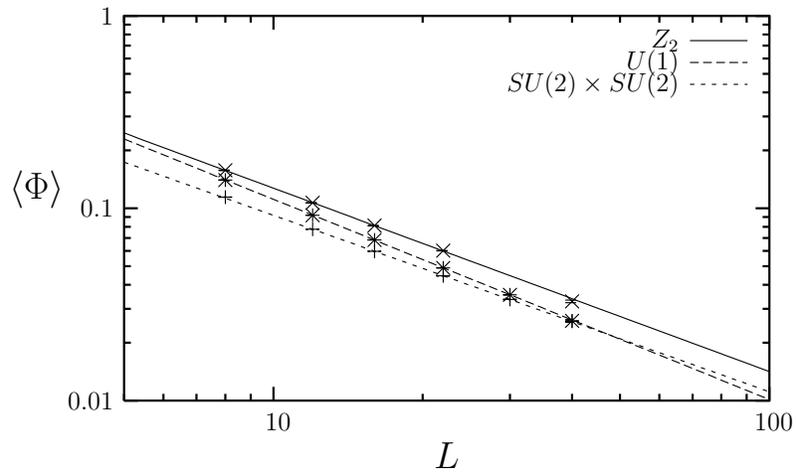, width=10.5cm}
\end{center}
\caption{Effective order parameter $\langle \Phi \rangle$ as a function of the lattice size $L$ for
all three models.}
\label{fig:mag}
\end{figure}
\begin{figure}[b!]
\bigskip\bigskip
\begin{center}
\epsfig{file=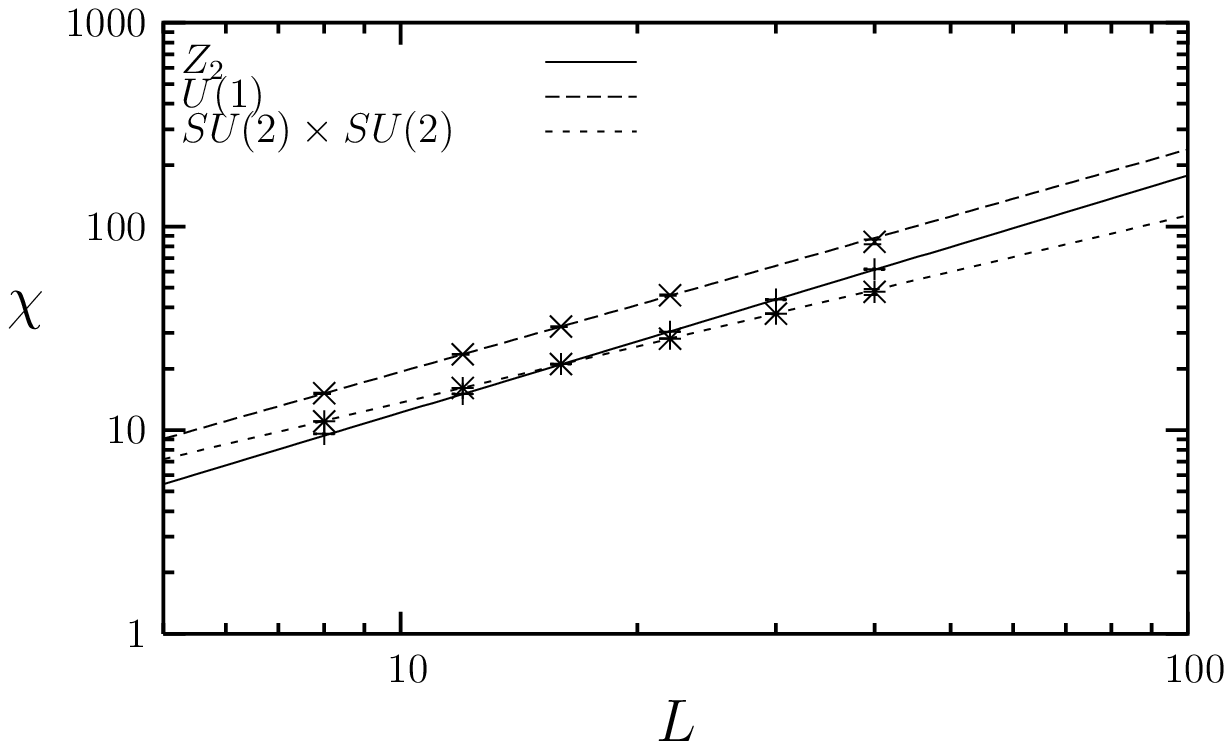, width=10.5cm}
\end{center}
\caption{Susceptibility $\chi$ as a function of the lattice size $L$ for all three models.}
\label{fig:susc}
\end{figure}
We performed the simulations on the largest $40^3$ lattice 
at a single value of the coupling $\beta=0.835$ and we generated approximately 
half a million configurations with average trajectory length equal to 1.0. We performed the simulations on the 
other lattices ($8^3, 12^3, 16^3, 22^3, 30^3$)
at all values $\beta=0.82, 0.83, 0.84, 0.85$ with approximately half to one million configurations for each $\beta$. 
It is clear that in the $Z_2$   model, the $U_B$ curves intersect at $(\beta_c, U_B(\beta_c)) = (0.835(1),0.232(8))$.

As expected, the situation is somewhat different in the $U(1)$  model.
In this model infrared fluctuations are stronger than in the discrete symmetry model. As a result, 
finite size effects near the critical coupling are larger for $U(1)$ than for $Z_2$.
We performed the simulations for the $U(1)$ model on the $40^3$ lattices 
at all values $\beta=0.830,0.835,0.840,0.845,0.850,0.86$, whereas on the smaller lattices  
at all values $\beta=0.83,...,0.86$  and in steps of $0.01$. 
The data set generated on $30^3$ at $\beta=0.850$ was 
corrupted and it was not included in the analysis. 
Approximately $6 \times 10^5$ - $1.3 \times 10^6$ configurations were generated at each $\beta$.
We show the values of $U_B$ versus $\beta$ in Fig.~\ref{fig:bind_u1}. 
The leading $L_1/L$ finite size corrections are taken into consideration by using eq.~(\ref{eq:confl}). 
We plot ($1/{\ln(L_1/L)}, 1/\beta_*$) for $L=8$ in Fig.~\ref{fig:bind_betac}. We computed the errors for $1/\beta_*$
from the jacknife errors for $U_B(\beta)$. The extrapolation of $\frac{1}{\beta_*(L)}$ to the point $1/\ln(L_1/L) =0$
gives $\beta_c=0.853(2)$ and $U_B(\beta_c)=0.424(8)$ on the $40^3$ lattice.

We performed an analysis for the $U_B$ data for $SU(2) \times SU(2)$ similar to the one for the $U(1)$  model. 
In this case, we performed
simulations at $\beta=0.92,...,1.00$ in steps of $0.02$ for the $8^3, 12^3, 16^3, 22^3, 30^3$ 
lattices and at $\beta=0.94,...,0.98$ in steps of $0.01$
on the largest $40^3$ lattice. 
The curves $U_B$ versus $\beta$ obtained from histogram reweighting at two consecutive values of $\beta$
did not intersect. Therefore, to obtain the intersection we used a linear approximation in the middle region between two curves.
The values of $U_B$ on different lattices 
near $\beta_c$ are shown in Fig.~\ref{fig:bind_su2} and the extrapolation of $1/\beta_*$ to the point 
$1/{\rm \ln}(L_1/L) =0$ are shown in Fig.~\ref{fig:bind_betac}. We extracted from this analysis the values $\beta_c=0.960(3)$ 
and $U(\beta_c)=0.544(7)$ on the largest $40^3$ lattice. 

Next, we calculated the exponent ratios $\beta_m/\nu$ for the three models 
by fitting to eq.~(\ref{eq:magn1}) the values of $\langle \Phi \rangle$ at $\beta_c$ obtained
on different lattice sizes. 
After fitting the data obtained on all lattice sizes we get $\beta_m/\nu=0.927(15)$ for $Z_2$, 
$\beta_m/\nu=0.955(20)$ for $U(1)$, and  
$\beta_m/\nu=1.04(2)$ for $SU(2) \times SU(2)$. These values of $\beta_m/\nu$ take into consideration the 
statistical error in $\beta_c$. In an effort to check to what extend our results are affected by possible small volume effects
we repeated the analysis without including the smallest lattice. Our results, summarized in table \ref{tab:t1} show that 
any finite volume systematic effects are smaller than the statistical errors. Another analysis where the $12^3$ data 
were excluded confirmed this conclusion.
The data and the fitted functions (for $L=12,...40$) for the three models are shown in Fig.~\ref{fig:mag}. 

Similarly, we obtained the exponent ratios $\gamma/\nu$ by fitting the
data for the susceptibility $\chi$ (eq.~(\ref{eq:chi1})) at $\beta_c$
to its FSS relation eq.~(\ref{eq:chi.fss}).
We present our results in Table \ref{tab:t2} together with analytical predictions
obtained from large-$N_f$ calcultations to order $1/N_f^2$ \cite{gracey.z2a,gracey.su2.u1}. 
It is clear that our numerical results are in good agreement with the analytical predictions. Furthermore, the 
results we got after omitting the smallest $8^3$ volume show that any finite size systematic
effects are within the statistical errors. The data and the fitted functions (for $L=12,...,40$) for the three models 
are shown in Fig.~\ref{fig:susc}. 

\begin{figure}[ht]
\bigskip\bigskip
\begin{center}
\epsfig{file=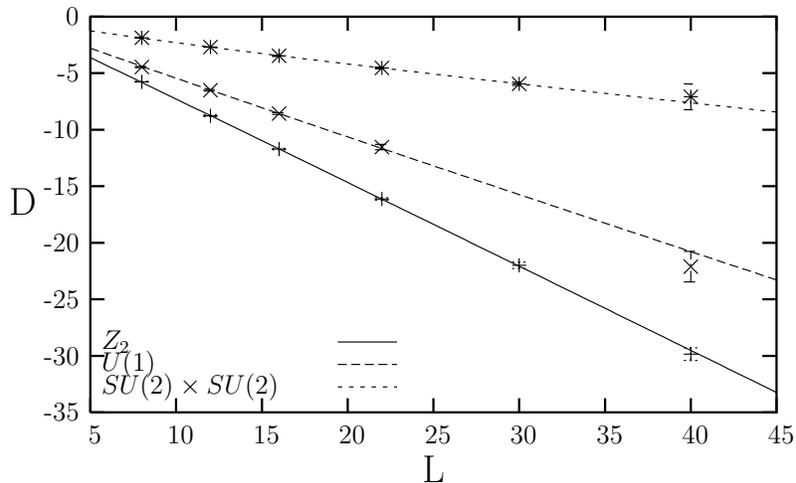, width=10.5cm}
\end{center}
\caption{Logarithmic derivarive $D$ of the order parameter as a function of the lattice size $L$ for all three models.}
\label{fig:logder}
\end{figure}

We used the logarithmic derivative $D$, defined in eq.~(\ref{eq:derlog}),
to calculate the exponent $\nu$.
According to eq.~(\ref{eq:ld}), at $\beta_c$, $D \sim L^{1/\nu}$. 
We present the values of $\nu$ for each model 
in Table \ref{tab:t3} together with the respective values obtained from large-$N_f$ calculations to order 
$1/N_f^2$ \cite{gracey.z2b, gracey.u1b, gracey.su2b}. 
As in the $\gamma/\nu$ case, the results obtained from our simulations are in good agreement with the analytical 
predictions. Like in the previous observables, in this case also systematic small volume effects are within the statistical 
errors. The data and the fitting functions (for $L=12,...,40$) for the three models at their critical couplings are shown 
in Fig.~\ref{fig:logder}.

Using our results for $\beta_m/\nu$ and $\gamma/\nu$, obtained from fits on all lattice sizes, 
we can check whether the hyperscaling relation
\begin{equation}
\frac{\beta_m}{\nu} + \frac{1}{2}\frac{\gamma}{\nu} - \frac{d}{2} = 0
\label{eq:hyper}
\end{equation}
is satisfied.
We find that for all three different models the left hand side of eq.~(\ref{eq:hyper}) is consistent with the value zero 
with a statistical uncertainty of $3$-$4 \%$. 

\begin{table}[tbp]
\centering \caption{Values of $\beta_m/\nu$ measured from our simulations.}
\medskip
\label{tab:t1}
\setlength{\tabcolsep}{1.0pc}
\begin{tabular}{cccc}
\hline \hline
                                 & $Z_2$          &   $U(1)$       &  $SU(2) \times SU(2)$  \\
\hline
simulations $L=8,...,40$         & 0.927(15)      &  0.955(20)     &   1.04(2)            \\
simulations $L=12,...,40$        & 0.917(20)      &  0.952(25)     &   1.05(3)            \\
\hline \hline
\end{tabular}
\end{table}

\begin{table}[tbp]
\centering \caption{Values of $\gamma/\nu$ measured from our simulations and from large-$N_f$ calculations.}
\medskip
\label{tab:t2}
\setlength{\tabcolsep}{1.0pc}
\begin{tabular}{cccc}
\hline \hline
                                        & $Z_2$          &   $U(1)$         &  $SU(2) \times SU(2)$  \\
\hline
simulations $L=8,...,40$                & 1.152(25)        & 1.09(3)        &   0.925(25)            \\
simulations $L=12,...,40$               &  1.165(40)       &  1.09(4)       &   0.910(30)            \\        
large $N_f$ \cite{gracey.z2a,gracey.su2.u1}   & 1.132           &  1.06          &    0.946           \\
\hline \hline
\end{tabular}
\end{table}

\begin{table}[tbp]
\centering \caption{Values $\nu$ measured from our simulations and from large-$N_f$ calculations.}
\medskip
\label{tab:t3}
\setlength{\tabcolsep}{1.0pc}
\begin{tabular}{cccc}
\hline \hline
                                                         & $Z_2$         &   $U(1)$     &  $SU(2) \times SU(2)$  \\
\hline
simulations $L=8,...,40$                                 &  0.98(2)       &  1.05(2)     &    1.14(3) \\
simulations $L=12,...,40$                                &  0.99(2)       &  1.03(4)     &    1.16(5)  \\             
large $N_f$ \cite{gracey.z2b, gracey.u1b, gracey.su2b}     &  0.98        &  1.02        &    1.11         \\
\hline \hline
\end{tabular}
\end{table}
                                                                                                                                 
\section{Conclusions}
We presented results from Monte Carlo simulations of $3d$ four-fermion models with $Z_2$, $U(1)$, 
and $SU(2) \times SU(2)$ chiral symmetries. These models are among the simplest relativistic field theories
of interacting fermions, and therefore are benchmarks for studying critical phenomena
in the presence of massless fermions. They are also used as model field theories to study 
the behavior of strong interaction under extreme conditions and have applications in condensed
matter systems.
In all three cases, we performed simulations with $N_f=4$ fermion species. Analytical 
calculations predict small next-to-leading order corrections for the critical exponents of the second order phase
transitions of these models at this intermediate value of $N_f$. We detected these corrections in our simulations 
by employing standard finite size scaling techniques and we found them to be in good 
agreement with large-$N_f$ expansions up to $O(1/N_f^2)$ 
\cite{gracey.z2a,gracey.z2b,gracey.su2.u1,gracey.u1b,gracey.su2b}.
Our results improve significantly previous numerical studies of $3d$ four-fermion models. Future work with 
much better statistics on a variety of lattices including larger sizes than the ones used in this work will allow for 
the detection of corrections to scaling effects and possible deviations
from the $O(1/N_f^2)$ analytical calculations. Also simulations with $N_f=1$ will be particularly instructive, as 
for such a small $N_f$ large-$N_f$ calculations cannot be applied. 
 
\section*{Acknowledgements}
The simulations were performed on a cluster of 64-bit AMD Opterons 250 at the Frederick Institute of
Technology, Cyprus. Discussions with John Gracey and Simon Hands are greatly appreciated.



\begin{thebibliography}{99}
\bibitem{rosenstein.1} B. Rosenstein, B.J. Warr, and S.H. Park, Phys. Rev. Lett. {\bf 62}, 1433 (1989).
\bibitem{justin} J. Zinn-Justin, Nucl. Phys. B {\bf 367} 105 (1991).
\bibitem{rosenstein.2} B. Rosenstein, B.J. Warr, and S.H. Park, Phys. Rep. {\bf 205}, 59 (1991). 
\bibitem{hands} S. Hands, A. Koci\'c, J.B. Kogut, An. Phys. {\bf 224}, 29 (1993). 
\bibitem{derkachov} S.E. Derkachov, N.A. Kivel, A.S. Stepanko, and A.N. Vasil'ev,
Theor. Math. Phys. {\bf 22}, 1047 (1992).
\bibitem{gracey.z2a} J.A. Gracey, Z. Phys. C {\bf 59}, 243 (1993).  
\bibitem{gracey.z2b} J.A. Gracey, Int. J. Mod. Phys. A {\bf 9}, 567 (1994).  
\bibitem{gracey.su2.u1} J.A. Gracey, Phys. Lett. B {\bf 308}, 65 (1993).  
\bibitem{gracey.u1b} J.A. Gracey, Phys. Rev. D {\bf 50}, 2840 (1994).  
\bibitem{gracey.su2b} J.A. Gracey, Z. Phys. C {\bf 61}, 115 (1994).  
\bibitem{wetterich.1} L. Rosa, P. Vitale, and C. Wetterich, Phys. Rev. Lett. {\bf 86}, 958 (2001).
\bibitem{wetterich.2} F. Hofling, C. Nowak, and C. Wetterich, Phys. Rev. D {\bf 66} 205111 (2002).
\bibitem{karkkainen} L. Karkkainen, R. Lacaze, P. Lacock, and B. Peterson, Nucl. Phys. B {\bf 415}, 781 (1994).
\bibitem{focht} E. Focht, J. Jersak, and J. Paul, Phys. Rev. D {\bf 53} 4616 (1996).  
\bibitem{general} K.G. Klimenko, Z. Phys. C {\bf 37}, 457 (1988);
B. Rosenstein, B.J. Warr and S.H. Park, Phys. Rev. D{\bf 39}, 3088 (1989);
S.J. Hands, A. Koci\'c and J.B. Kogut, Nucl. Phys. B{\bf 390}, 355 (1993);
J.B. Kogut, M.A. Stephanov and C.G. Strouthos, Phys. Rev. D{\bf 58}:096001 (1998);
S. Chandrasekharan, J. Cox, K. Holland and U.J. Wiese, Nucl. Phys. B {\bf 576} 481 (2000);
C.G. Strouthos and S. Christofi, JHEP {\bf 0501}:057 2005.
J.B. Kogut and C.G. Strouthos, Phys. Rev. D{\bf 63}:054502 (2001);
S.J. Hands, B. Lucini and S.E. Morrison, Phys. Rev. D{\bf 65}:036004 (2002).
\bibitem{sharkar} R. Sharkar, Phys. Rev. Lett. {\bf 63}, 203 (1989);
N. Dorey and N.E. Mavromatos, Phys. Lett. B {\bf 250}, 107 (1990); Nucl. Phys. B {\bf 386}, 614 (1992).
\bibitem{aitchison} I.J.R. Aitchison and N.E. Mavromatos, Phys. Rev. B{\bf 53}, 9321 (1996).
\bibitem{herbut} I. Herbut, Phys. Rev. Lett. {\bf 97}, 146401 (2006).
\bibitem{wang} S.J. Hands, A. Koci\'c, J.B. Kogut, R.L. Renken, D.K. Sinclair, and K.C. Wang, Nucl. Phys. B {\bf 413}, 503 (1994).
\bibitem{hands.u1} S.J. Hands, S. Kim and J.B. Kogut, Nucl. Phys. B {\bf 442}, 364 (1995).
\bibitem{hands.su2} S.J. Hands and S.E. Morrison, Phys. Rev. D{\bf 59}, 116002 (1999).
\bibitem{barber} M.N. Barber, in {\it Phase Transitions and Critical Phenomena}, edited by C. Domb and J. Lebowitz 
(Academic, New York, 1983).
\bibitem{binder} K. Binder, Z. Phys. B{\bf 43}, 119 (1981). 
\bibitem{holm} C. Holm and W. Janke, Phys. Rev. B{\bf 48}, 936 (1993).
\bibitem{histo} A.M. Ferrenberg and R.H. Swendsen, Phys. Rev. Lett.  {\bf 61}, 2635 (1988).
\end{thebibliography}
\end{document}